\documentclass[a4paper,aip,jcp,floatfix,unsortedaddress,superscriptaddress,reprint]{revtex4-1}
\usepackage{graphicx}

\begin{document}

\title{A comparison between parallelization approaches in molecular dynamics simulations on GPUs}
\author{Lorenzo Rovigatti}
\email[Author for correspondence, ]{lorenzo.rovigatti@uniroma1.it}
\affiliation{Dipartimento di Fisica, Sapienza---Universit\`a di Roma, Piazzale
A. Moro 5, 00185 Roma, Italy}
\affiliation{Faculty of Physics, University of Vienna, Boltzmanngasse 5, A-1090 Vienna, Austria}
\author{Petr \v{S}ulc}
\affiliation{Rudolf Peierls Centre for Theoretical Physics, University of
Oxford, 1 Keble Road, Oxford, OX1 3NP, United Kingdom}
\author{Istv\'an Z. Reguly}
\affiliation{Oxford e-Research Centre, University of Oxford, Oxford,
7 Keble Road, Oxford, OX1 3QG, United Kingdom}
\author{Flavio Romano}
\affiliation{Physical \& Theoretical Chemistry Laboratory, Department of
Chemistry, University of Oxford, South Parks Road, Oxford, OX1 3QZ, United
Kingdom}

\date{\today}

\begin{abstract}
We test the relative performances of two different approaches to the computation of
forces for molecular dynamics simulations on Graphics Processing Units. A
``vertex-based'' approach, where a computing thread is started per particle, is
compared to an ``edge-based'' approach, where a thread is started per each
potentially non-zero interaction. We find that the former is more efficient for
systems with many simple interactions per particle, while the latter is more
efficient if the system has more complicated interactions or fewer of them. By
comparing computation times on more and less recent GPU technology, we predict
that, if the current trend of increasing the number of processing cores -- as
opposed to their computing power -- remains, the ``edge-based'' approach will
gradually become the most efficient choice in an increasing number of cases.
\end{abstract}
 
\maketitle

\section{Introduction}

For a long time the steady increase in frequency of CPUs guaranteed an
almost equally steady increase in the performance of compute-intensive
applications with very little effort on the programming side. However, by 2005,
it became clear that, essentially due to power dissipation constraints, this trend
would not last for much longer. In order to keep increasing the computational
capacity, multi-core CPUs have appeared, the size of on-chip caches has grown
and the complexity of control circuitry has increased. At the same time, new
architectures have emerged; Graphical Processing Units (GPUs), originally
special-purpose hardware for graphics, were equipped to carry out general
purpose computations. While most CPU architectures are still aimed at
low-latency handling of a few processes, GPUs support a massive amount
of parallelism, giving up low latency in exchange for high throughput. By
integrating up to thousands of simple computing units, grouping them under just
a few dozen control units and connecting them to high-speed memory, they are
capable of supporting up to tens or hundreds of thousands of threads executing
concurrently. This enables GPUs to deliver an order of magnitude higher
performance than CPUs on workloads with a high amount of data parallelism, i.e.
where the same operations are carried out over different data. Efficient use of
this hardware requires employing parallel programming models that are very
explicit about data parallelism and often expose low-level features of the GPU.
The situation is further complicated by significant architectural changes when
a new generation of GPUs is released: improved hardware and new features may affect the 
relative performance of different algorithms, often permitting some of them to run
more efficiently and slowing down others.

Computer simulations are a very valuable tool in many areas of
science.~\cite{feynman,waterman1995introduction,jensen2007introduction,frenkelsmith,allen1989computer,LandauBinderBook}
The relatively cheap computer power offered by GPUs is very attractive to
simulators, since it allows extended time scales and large system sizes to be
investigated.\cite{owens2008gpu,nickolls2010gpu,Stone2010,van2008harvesting}
But while some simulation algorithms are relatively easy to parallelize, some
others are very difficult to code efficiently on a parallel machine. Molecular
dynamics (MD), in its many variants, is a prominent technique in computational
physics and chemistry, and it is in principle an algorithm that is suitable for
parallelization.\cite{frenkelsmith,allen1989computer,Stone2010,plimpton1995fast}
Most free and commercial simulation packages have the option to run in
parallel, and a growing number of them offers the option to run on
GPUs.\cite{hoomd,brown2011implementing,pronk2013gromacs,namd,openmm}
The different simulation packages exploit GPUs in different ways, each using a
distinct approach to overcome the two main obstacles to make full use of a GPU:
potentially concurrent writes to the same memory location and having a large
number of balanced tasks for the GPU to carry out.

An important difference between CPUs and GPUs is that the latter have a much
smaller cache, and therefore the optimisation of memory access patterns has
received a lot of attention; since many scientific computations are bound by
the amount of data they have to move, it is crucial to achieve as high a
bandwidth as possible.

GROMACS~4.6, for example, uses the GPUs for the calculation of non-bonded
interactions\cite{gromacs_gpu_algorithm} and a parallel reduction algorithm to
add the calculated forces between pairs of particles, rearranging particles in
memory to speed-up memory access. NAMD\cite{namd} also implements a GPU-based
calculation of non-bonded forces,\cite{stone2007accelerating} where forces for
each pair of interacting particles are calculated twice. HOOMD-blue\cite{hoomd} and
LAMMPS\cite{brown2011implementing} also compute forces for each interacting
pair twice to avoid atomic operations or memory synchronization bottlenecks.

Different parallelization approaches of the MD algorithm have been considered
for various platforms. Three main parallel decomposition schemes typically
considered are:\cite{plimpton1995fast}
\begin{itemize}
 \item Atom-decomposition. The particles in the simulation are split between the computing units, which calculate the interaction forces between them. 
 \item Force-decomposition. The interactions between particles are split in a way that each computing unit is responsible for a particular subset of forces between particles to be evaluated. 
 \item Space-decomposition. The domain of the simulated system is partitioned into smaller subsets, with each computing unit responsible for calculation of forces and updates of positions of particles in its assigned subdomain. 
\end{itemize}

Each of the proposed schemes requires communication between the computing
units. The choice of the optimal parallelization scheme hence depends on the
system studied as well as on the properties of the hardware architecture. The
communication cost and the load balance between computing units then determine
the optimal parallelization scheme.

In this work, we implement and study the atom-decomposition approach (which we
call the ``vertex-based'' approach) and a variant of the force-decomposition
approach (which we call the ``edge-based'' approach) on GPUs, where one
computing unit corresponds to a single thread of the graphics card that will be
carrying out the computations. In the ``vertex-based'' approach, a computing
thread is started per each particle and the force ${\mathbf F}_{ij}$
between each pair of particles $i$ and $j$ is calculated separately for $i$ and
$j$. This might seem like a waste of computational resources, since it is known
by Newton's third law that $\mathbf{F}_{ij}=-\mathbf{F}_{ji}$, but it has the
advantage of being more parallelizable. In
the Supplementary Information we show that often the work saved by using
Newton's third law is outbalanced by the fact that the force calculation is less
parallel, and in all cases the performances of the two approaches are similar.
In the ``edge-based'' approach, a thread is started per each potentially
non-zero interaction, and we use atomic operations and Newton's third law,
$\mathbf{F}_{ij} = -\mathbf{F}_{ji}$, to calculate the resulting force acting
on each particle. 

We study the two parallelization approaches by implementing a GPU-based MD
algorithm for three different model molecules (Lennard-Jones particles, patchy
particles and coarse-grained DNA, see Section~\ref{sec_models}) that differ
substantially in the complexity and physical features of the interaction
potential and we report the relative performance of the algorithms for
different hardware.

To our best knowledge, the edge-based approach has not yet been systematically
compared to the vertex-based approach on GPU. Most GPU implementations use the
atom-decomposition (i.e.,~vertex-based in our terminology)
scheme.\cite{liu2008accelerating,stone2007accelerating,van2008harvesting,hoomd,brown2011implementing}
Zhmurov and collaborators considered both atom and force decomposition schemes
for an MD study of a coarse-grained polymer model on GPU~\cite{zhmurov2010sop},
but a comparison of the performance of the two approaches was not reported.

Interestingly, we find that the edge-based approach holds better performance on
newer hardware for short-ranged, anisotropically interacting systems. This
broad class of coarse-grained models has recently gained significant attention 
from the soft matter and biophysics community as a tool for the investigation 
of biological macromolecules\cite{depablo2,dna_pccp} and self-assembling
processes~\cite{RapaportVirusPRL08,self_knotting_patchy} and for the synthesis
of new materials~\cite{Romano12b,dna_tetra_gel}. The performance boost provided
by the edge-approach, which can be readily implemented in any modern MD
package, will help in exploiting the power of the GPUs in this field with even
more efficiency.

\section{Methods}
\label{sec_methods}
For our performance tests, we perform Brownian dynamics simulations in the
$NVT$ ensemble, with the thermostat described in Ref.~\onlinecite{John_valence}. 
Thermostating the system with this method does not significantly affect performances, 
and thus we do not expect this thermostat to bias our comparison in any way.
We implement a combination of fairly standard
optimizations in order to speed-up performance. We use Verlet
lists\cite{frenkelsmith} to provide each particle $i$ with a list $L_i$ of
all the particles inside a sphere centred on $i$ of radius $r_v = r_c + r_s$,
where $r_c$ is the interaction cut-off and $r_s$ is the Verlet skin. Verlet lists
are then updated by using a standard cell algorithm. On GPUs, both cell filling
and Verlet lists updating are performed on a one-particle-per-thread basis.
We perform simulations at different numbers of particles $N$ (nucleotides in the case of the DNA model).

In the case of the Lennard-Jones and patchy models (see
Section~\ref{sec_models} for details), we maximise cache hits by periodically
sorting particle data, stored in the global memory, according to a 8-vertex
Hilbert curve.\cite{hoomd} The resulting speed-up depends on $N$, ranging from 
$20\%$ to $100\%$ for $N > 10^4$ as discussed in
Ref.~\onlinecite{tesi_lorenzo}. We do not apply this procedure to DNA
simulations, since doing so does not result in any significant measurable 
gain.

One of the potential drawbacks of using GPUs is in the accuracy of floating
point operations. Even though double precision support is quickly improving,
the GPUs' peak double precision performance is only a half or a third of its
peak single precision performance, and it is thus crucial to use
single-precision performance as much as possible. Unfortunately, it has been
shown that lengthy single precision simulations lack reliability even for
simple potentials.\cite{Colberg_gpu} In order to maximise performances and
minimise numerical instabilities, we use double precision calculations to carry
out the integration of positions and momenta and single precision calculations
to compute forces, as commonly done in many MD packages. This \textit{mixed precision} 
algorithm results in a performance decrease ranging between 10\% and 40\% compared 
to single precision (depending on the model and simulation parameters), but 
dramatically improves the numerical stability.\cite{tesi_lorenzo}

\begin{figure*}
\begin{center}
\includegraphics[width=0.9\textwidth]{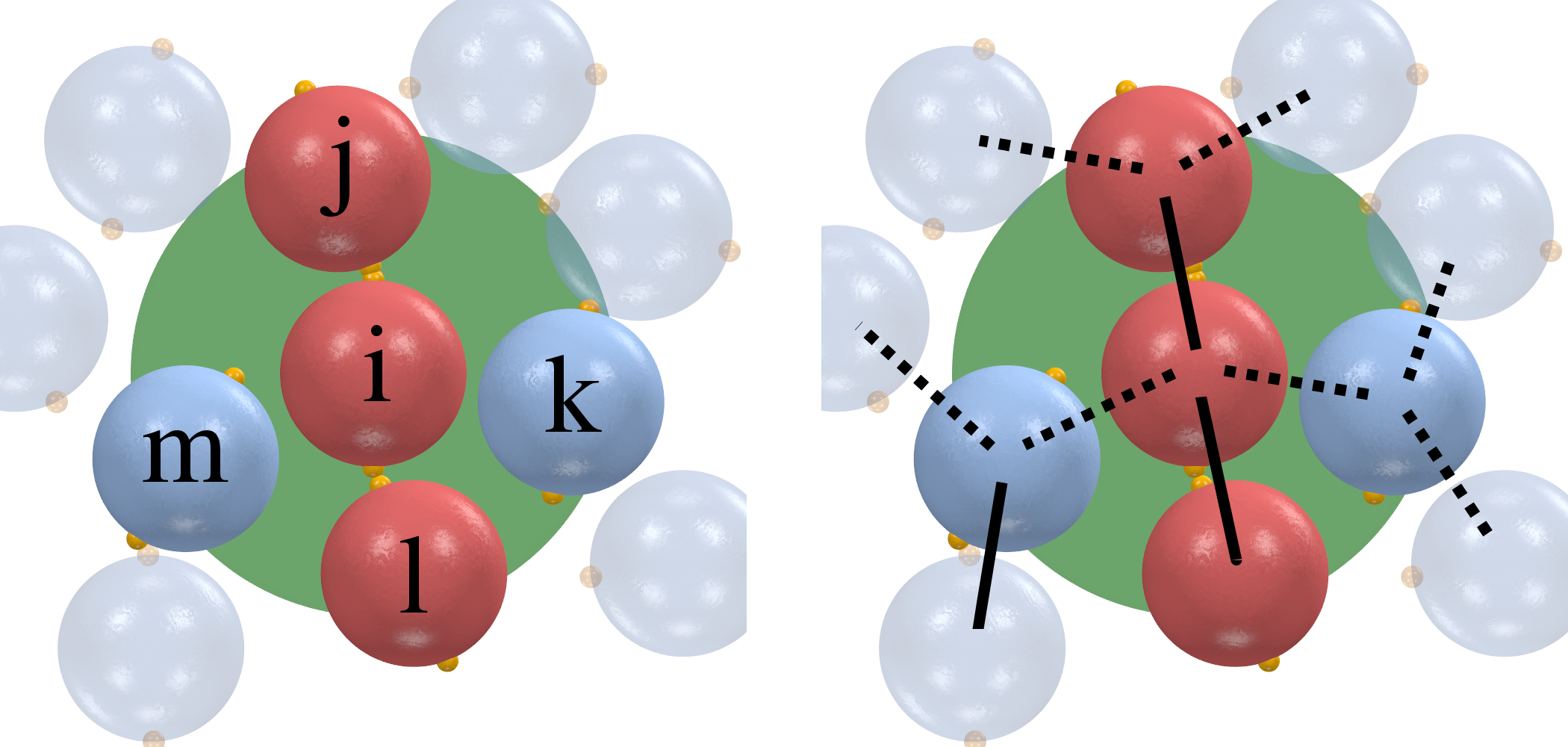}
\end{center}
\caption{\label{fig:cartoon} A cartoon showing how the (left) vertex and (right) edge 
algorithms work in a patchy system. Patches are depicted as yellow spheres. The
Verlet list of the central particle $i$ contains all the particles which are within a sphere
of radius $r_v$ (in green), $j$, $k$, $l$ and $m$. $i$ and the only two particles exerting a non-zero
interaction on it are colour-coded in red, while all the other particles are coloured in light blue. 
Particles whose centres are outside the Verlet sphere are semitransparent. In the vertex approach (left panel), the 
$i$-th thread computes the interaction between $i$ and each of the particles in its Verlet list. Thus, 
if we consider only particles within the green sphere, $5$ threads are started. By contrast, 
in the edge approach (right panel) a thread is started for each potential interaction between 
two particles, i.e. one for each of the lines connecting particle pairs in the figure. If, 
once again, we consider only particles within the green sphere, the total number of started 
threads is $10$, doubling the number of threads of the vertex case and thus enhancing parallelism. 
The only threads that require to perform expensive atomic operations to update the forces 
acting on the particles are those for which these forces are different from zero (full lines). 
In anisotropic models such as the patchy one presented here, a large part of the potential 
interactions is effectively zero (dashed lines) and hence does not result in any concurrent 
memory write.
}
\end{figure*}

We now describe in more detail the two parallelization approaches used in this
work. A cartoon providing a visual explanation of the two algorithms is shown in~\ref{fig:cartoon}.
The first one, which we call ``vertex-based'', is
to start a thread for each of the particles in the system. This thread will go
through a list of potentially interacting neighbours, compute the force coming
from each of the neighbours, and then add them together to yield the total
force acting on the particle. The threads done in this way are completely
independent, since they do not need to write concurrently to the same memory. 

The second parallelization approach we consider is a more aggressively parallel
``edge-based'' approach, where a thread is started for each potentially
non-zero interaction (or equivalently for each potentially interacting pair of
particles). Therefore, the total number of threads is equal to the sum of the
number of neighbours of each particle divided by two. Each thread will compute
the force due to a specific pair interaction and add it to the total force
acting on both particles $i$ and $j$. However, multiple threads trying to
concurrently apply a force to the same particle pose a data race which has to
be resolved: the most general approach is for the force computation to store
the three components of force vectors on a per-edge basis, and once all the
threads have finished, carry out a parallel segmented reduction to add up the
contributions from each interaction for each particle. The second approach is
to use atomic operations to directly accumulate the total force acting on a
particle. However, compared to regular memory transactions, atomic operations
are very expensive. Therefore the update should not be performed naively, but
only if the force is actually non-zero. This is a common occurrence when
treating anisotropic interactions, since it is possible to have particles
separated by a distance $r<r_c$ but mutually oriented in such a way that the
force acting between them is zero, but is also relatively common when treating
isotropic interactions since the Verlet lists always contain a significant
fraction of non-interacting particles. Threads computing an interaction
between two particles that turns out to be zero thus do not carry out any
atomic operations. We stress that in this approach it is natural to exploit
Newton's third law, since many different threads compute the forces acting on a
single particle and thus threads cannot be independent by design.

The vertex-based approach has the advantage of being ``embarrassingly
parallel''; threads do not have to communicate or synchronize in any way. On
the other hand, the amount of threads that get started at the same time is
equal to the number of particles in the system, which poses a lower bound on
the computer time required per step: if there are fewer particles than the
number of threads required to saturate the GPU (we point out that the current
trend is to increase the number of concurrent threads), there will be no
computational benefit in studying smaller systems using the vertex-based
approach. We also point out that the edge-based approach is only effective if
treating systems with short-range forces (i.e., forces that vanish faster than
$r^{-3}$ in $3D$), since otherwise $\mathcal{O}(N^2)$ threads would need to be
started. While this
approach is in principle feasible for small values of $N$, the many concurrent
updates to the force vector acting on each particle would make it not
competitive.

It is important to identify the bottlenecks in each of the approaches, since
the convenience in using one parallelization approach or the other will depend
on how each particular case is affected by these bottlenecks. The vertex-based
approach is limited by insufficient parallelism at low $N$, and by the
computational time required by the slowest thread in each of the warps, and
thus by the computation of the forces of the particle that has the most and/or
most complicated interactions. The edge-based approach, on the other hand, is
limited by the most expensive pair interaction, a much lower bound, and by the
throughput of atomic operations.

It may be worth addressing in more detail whether it would be worth 
to exploit Newton's third law
($\mathbf{F}_{ij} = -\mathbf{F}_{ji}$) in the vertex-based approach. As
discussed earlier, this would half the total number of computations. However,
additional operations are then necessary to update the total force acting on
the chosen particle. For this reason, many MD packages on GPUs prefer to repeat
the calculation to avoid concurrent writes, which are known to be a potential
source of
slowdown.\cite{liu2008accelerating,van2008harvesting,zhmurov2010sop,stone2007accelerating,hoomd}
To provide a quantitative analysis of this point, we also implemented a version
of the vertex-based approach that uses Newton's third law. Atomic addition is used
to update the total force acting on a particle. In order to avoid most of the
slowdown related to the atomic addition of the forces, each particle has $m$
force vectors on which the atomic adds are performed separately, and then the
vectors are combined to obtain the total force acting on each particle. The performances
turned out to be rather insensitive on the value of $m$, as long as $m > 10$.
We found that for
some of the studied systems the vertex based-approach with Newton's third law
was marginally (at most $1.2$ times) faster, and for the remaining cases
considered its performance was the same or worse than for the vertex-based
approach which did not exploit Newton's third law. Importantly, in all cases
where the vertex-based approach with Newton's third law is the fastest, it is
always slower than the edge-based approach. We hence consider the vertex-based
approach without the Newton's third law here and provide benchmarks comparing
the vertex-based approaches with and without Newton's third law in the Figure S2 in the
Supplementary Information.\cite{support}

All our simulations have been carried out on CUDA-enabled NVIDIA GPUs with
oxDNA, a simulation software originally developed to simulate a coarse-grained
DNA model,\cite{ouldridge_jcp,sulc_seq_dep} now extended to support
additional interaction potentials. The code is open source and can be freely
downloaded from the oxDNA website.\cite{dnawebsite} 

\begin{figure}
  \begin{center}
    \includegraphics[width=0.45\textwidth]{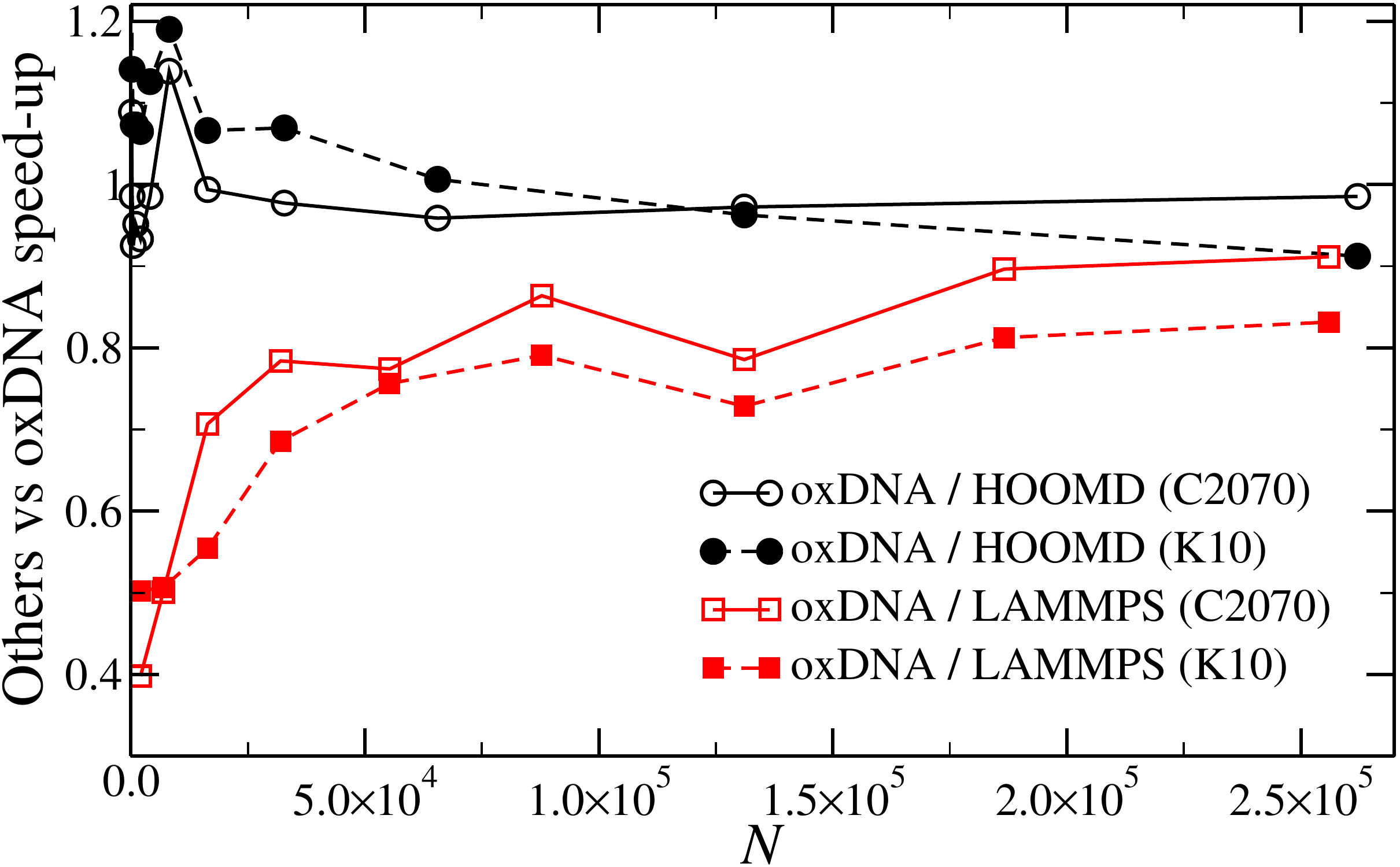}\\
  \end{center}
  \caption{\label{fig:benchmarks}
  Relative performance of oxDNA vs LAMMPS (in red) and HOOMD-blue (in black) for 
  different system  sizes on Tesla C2070 (full lines, empty symbols) and Kepler 
  K10 (dashed lines, full symbols) GPUs. A y-value larger (smaller) than one means 
  that oxDNA is slower (faster) than the other package.}
\end{figure}

It is important to ensure that our algorithm comparisons are carried out with a
simulation code of good overall efficiency, otherwise the results would be of
little use. We thus compared the oxDNA implementation of vertex-based
parallelization for Lennard-Jones (LJ) systems, a very common test potential,
consisting of up to 262\,144 particles with two popular simulation packages
that also allow for MD simulations on GPUs: LAMMPS\cite{brown2011implementing}
(1Feb14 version) and HOOMD-blue (0.11-3 version).\cite{hoomd} Figure~\ref{fig:benchmarks} 
shows some benchmarks, obtained on Tesla C2070 and Kepler K10 GPU cards. The speed 
of our code is within $10\%$ of the performance of the HOOMD-blue package for systems 
composed of 10000 particles or more, being slightly faster for larger systems. In
comparison with LAMMPS, our code is at least $20\%$ faster for all the systems considered.
More details can be found in the Supplementary Information (Fig.~S1)\cite{support},
which also contains a performance comparison between oxDNA and GROMACS 5.0.1 on a LJ
system showing that oxDNA is much faster than GROMACS in this case. 
We thus confirm that our implementation of the MD algorithm has performances 
comparable with two widely used parallel MD simulation packages oriented towards the soft 
matter community.

We note that GPU vs CPU relative performance depends dramatically on the hardware and 
even more on the simulation parameters and system under study. Common speed-ups for LJ 
and patchy systems of a single GPU vs a single CPU-core for large systems varies between 
factors of 20 and 100, depending mainly on the density.$^{12,29,41}$ As for the DNA model, 
the performance gain is usually between 20 and 50.~\cite{tesi_lorenzo}

\section{Models}
\label{sec_models}
In order to compare the performance of the two parallelization approaches on
diverse situations, we have implemented three different models with
substantially different features and different scopes. The first model is the
widely employed Lennard-Jones interaction potential, used to model
atomic and molecular systems such as noble gases and glass-forming
materials.\cite{frenkelsmith,kob_andersen} This iconic model potential, often
used to test algorithms, is a pairwise, spherically symmetric potential that
can be written in terms of the relative distance $r$ between two particles as

\begin{equation}
V(r) = 4\epsilon \left[ \left( \frac{\sigma}{r} \right)^{12} - \left( \frac{\sigma}{r} \right)^6 \right]
\end{equation}
where $\epsilon$ controls the depth of the attraction and $\sigma$ is the
particle diameter. As commonly done, we cut and shift the potential at a
distance $r_c = 2.5\sigma$.\cite{frenkelsmith}

The second interaction potential we implement, the patchy model, depends on the
relative orientations of each pair of particles as well as their relative
distance. Anisotropically interacting systems such as this one are becoming
increasingly popular in the soft matter field, due to the richness of phenomena
they exhibit\cite{bianchi2011patchy} and to the possibility of synthesising
particles with tunable shape and surface patterns.\cite{Glotz_Solomon_natmat}
Patchy particles are spherical colloids, i.e. nano-- or micro--sized particles,
with interacting spots decorating their surface. Lately, patchy particles have
been the subject of several theoretical, numerical and experimental studies,
and have been shown to exhibit novel and unexpected behaviour such as the
formation of stress-yielding, density-vanishing equilibrium
gels,\cite{John_valence,rovigatti_dos} crystallisation into open
lattices\cite{granicknature,Romano12b} or re-entrant gas-liquid phase
separations.\cite{Russo11a,reinhardt}
More specifically, we use an interaction potential that comprises a spherical hard-core-like repulsion
and a short-ranged interaction that depends on the relative orientations of
each pair of particles:

\begin{equation}
V(1,2) = V_{CM}(1,2) + V_P(1,2)
\end{equation}
where $V_{CM}$ is the interaction between the centres of mass
and $V_P$ is the interaction between the patches, modelled
as follows:

\begin{eqnarray}
 V_{CM}(12) & = & {\left( \frac{\sigma}{r_{12}}\right)}^{200} \label{eqn:potential_cm} \\
 V_{P}(12) & = & -\sum_{i=1}^{M}\sum_{j=1}^{M}\epsilon \label{eqn:potential_attractive}
\exp \left[ \frac{1}{2}{\left( \frac{r_{12}^{ij}}{0.12\sigma} \right)}^{n} \right]
\end{eqnarray}
where $r_{12}$ is the distance between the centres of mass, $r_{12}^{ij}$ is
the distance between patch $i$ on particle $1$ and patch $j$ on particle $2$
and $M$ is the number of patches per particle, which we fix to the value $M =
2$. This potential has been used in the past to study the dynamics of
patchy particles by means of MD simulations.\cite{John_valence,rovigatti_molphys}

The third and last model we use is oxDNA, a coarse-grained model specifically
designed to reproduce the mechanical, structural and thermodynamic properties
of DNA targeted to simulating processes occurring in DNA
nanotechnology.~\cite{tesi_tom,ouldridge_jcp,sulc_seq_dep} Indeed, oxDNA has
been used to investigate DNA nanotweezers,\cite{ouldridge_prl} DNA
walkers,\cite{ouldridge_walker}, a burnt-bridges DNA motor~\cite{sulc_burnt} and
other DNA motifs common in DNA nanotechnology.\cite{dna_pccp} By exploiting
GPUs, oxDNA can be used to investigate systems composed of thousands of
nucleotides.~\cite{tetramers_rovigatti,dna_tetra_gel} The basic
unit of the oxDNA model is a nucleotide, which is modelled as a rigid body
interacting with other nucleotides through a short-ranged, highly anisotropic
potential that takes into account contributions due to excluded volume,
backbone, stacking, coaxial stacking, cross-stacking and hydrogen bonding
interactions. The detailed form of the potential can be found in
Refs.~\onlinecite{ouldridge_jcp} and~\onlinecite{tesi_tom}.

From the point of view of the computational complexity, these three models have
different properties. In the LJ model, which is spherically symmetric and
relatively long-ranged ($r_{\rm c} = 2.5\sigma$), particles can have a large
number of interacting neighbours. The patchy model, on the other hand, is very
short-ranged ($r_c = 0.18\sigma$ in our case), and thus each particle has a
small number of potentially interacting neighbours, and on top of that its
anisotropic nature makes it so that each particle has an interaction which is
non-zero with only a fraction of its neighbours. Differently from the other two
models, oxDNA features a very complicated and computationally demanding
potential that, due to the large amount of branching and imbalance in the calculations 
required by particle pairs in different local environments, hinders performances 
on GPUs. Indeed, the GPU vs CPU speed-up for oxDNA is usually smaller than what we 
find for the LJ or patchy model. Similarly to the patchy model, the high degree of 
anisotropy of the potential results in a small number of interacting neighbouring particles.

The LJ and patchy models are simulated at three different values of the number
density, namely $\rho\sigma^3 = 0.1$, $0.3$ and $0.5$
The simulation temperature is $k_B T/\epsilon = 1.8$ for the LJ model and $k_B
T/\epsilon = 0.15$ for the patchy model. For the DNA model we perform
simulations of double-stranded octamers at a concentration of $2.7$ mM and at a
temperature of 300 $K$.

\section{Results}

\begin{figure}
  \begin{center}
    \includegraphics[width=0.45\textwidth]{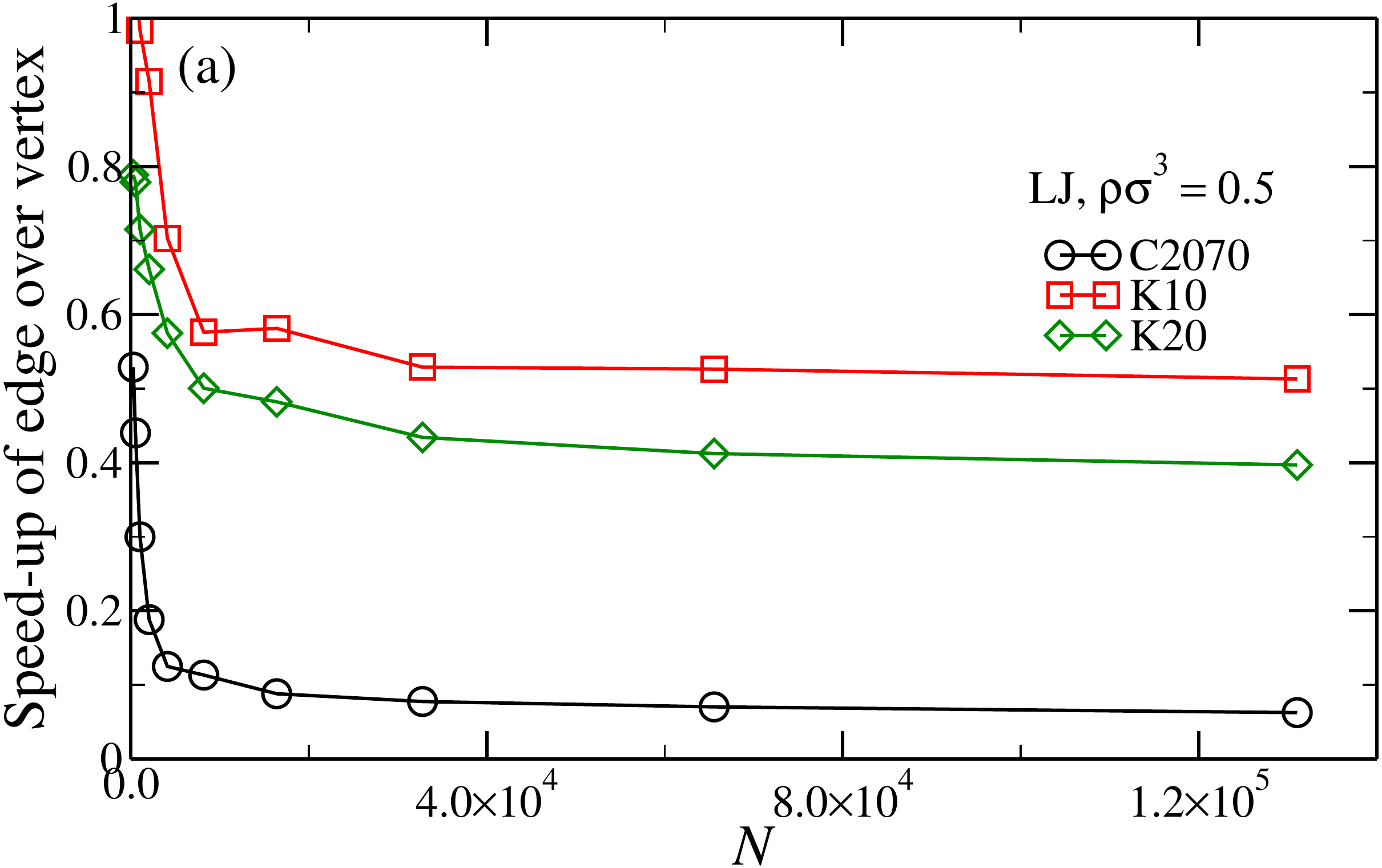} \\ ~\\
    \includegraphics[width=0.45\textwidth]{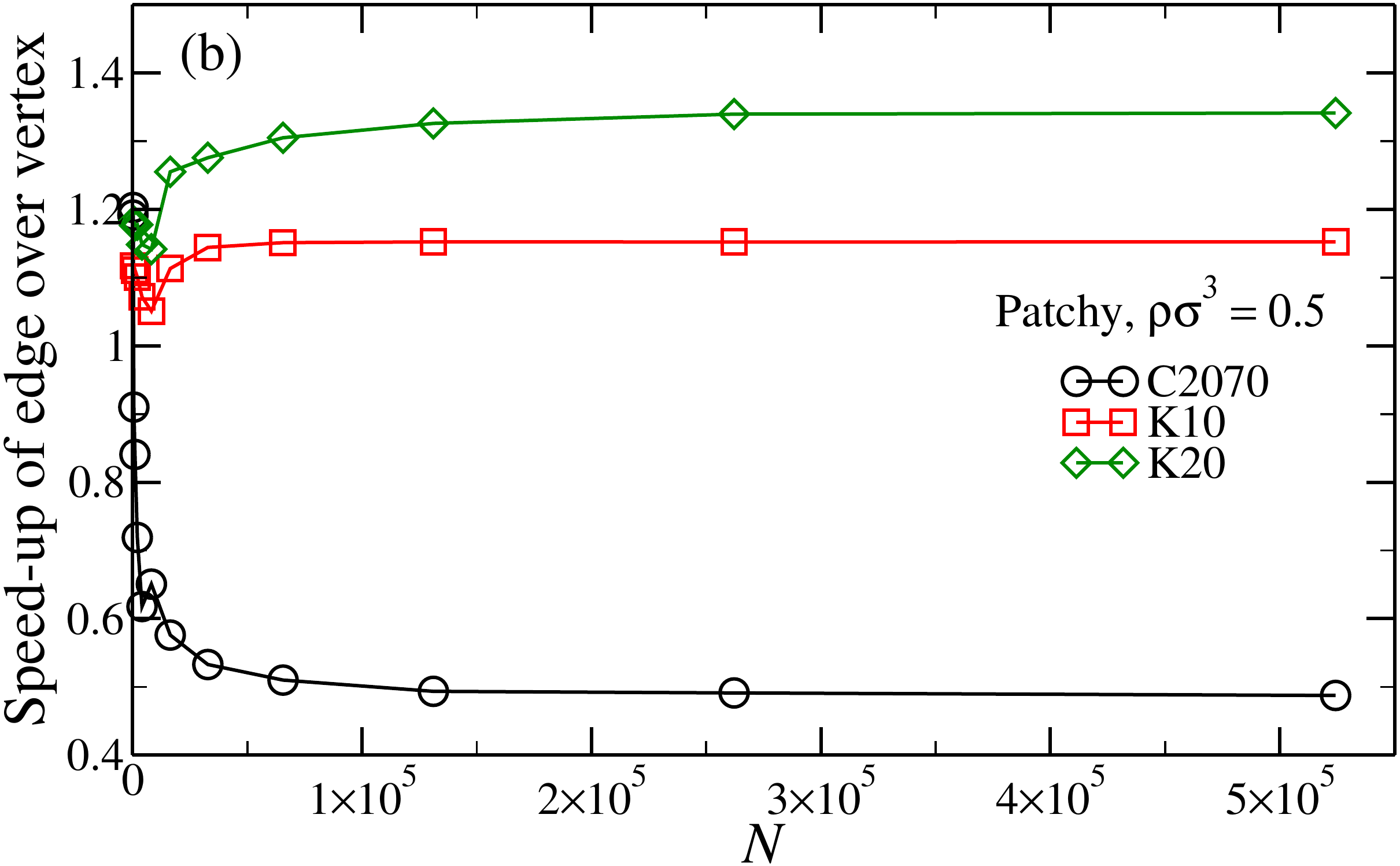}\\ ~\\
    \includegraphics[width=0.45\textwidth]{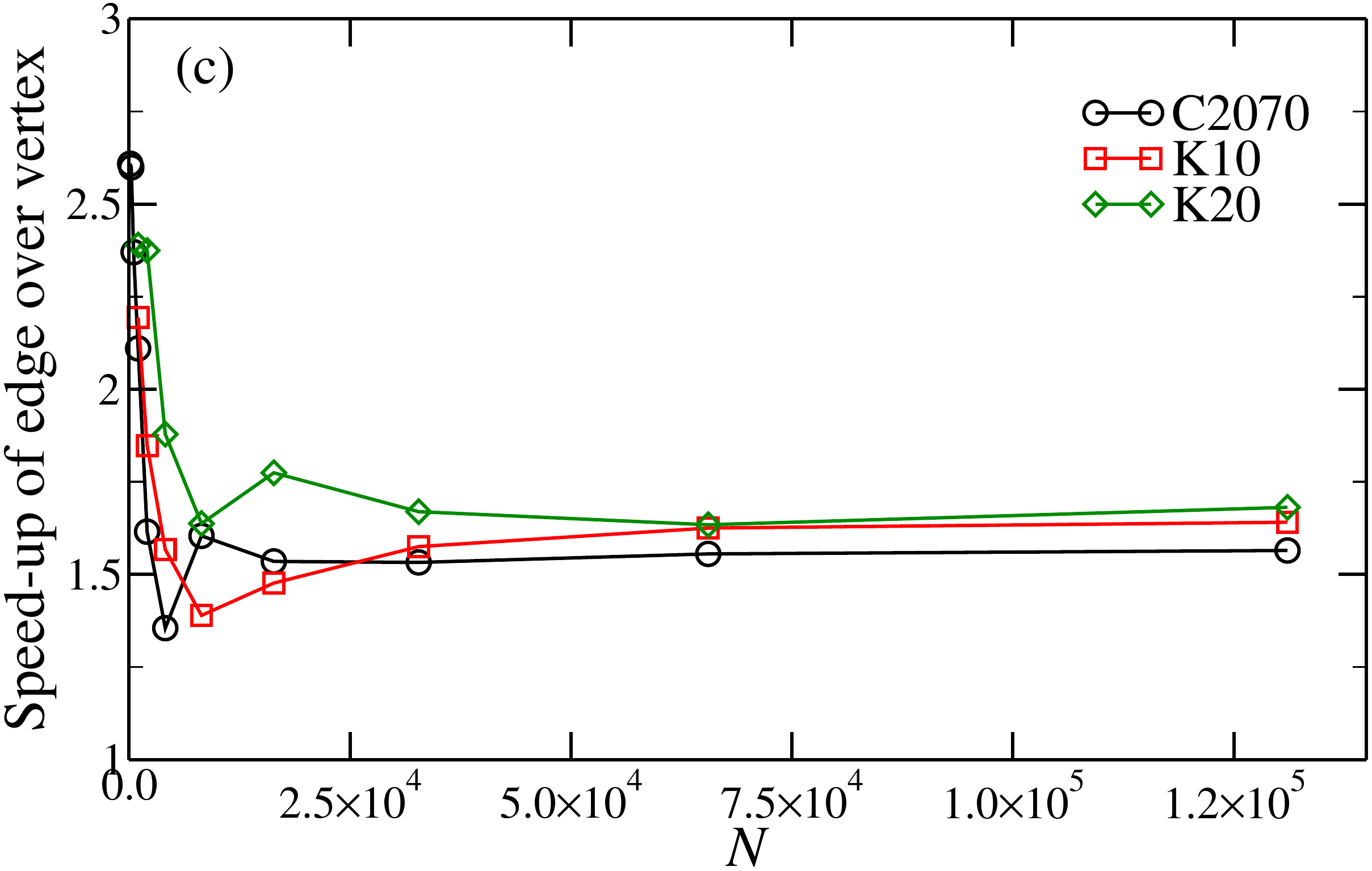}
  \end{center}
  \caption{\label{fig:timevsN}
Achieved speed-up for the edge approach over the vertex approach for (a) the LJ
model and (b) the patchy model at $\rho\sigma^3 = 0.5$ and (c) for the oxDNA model as a function
of the number of particles $N$ and for different GPUs. All bumps and spikes are reproducible. 
The edge (vertex) approach is more efficient when y-values are larger (smaller) than one.
}
\end{figure}

\begin{figure}
  \begin{center}
    \includegraphics[width=0.45\textwidth]{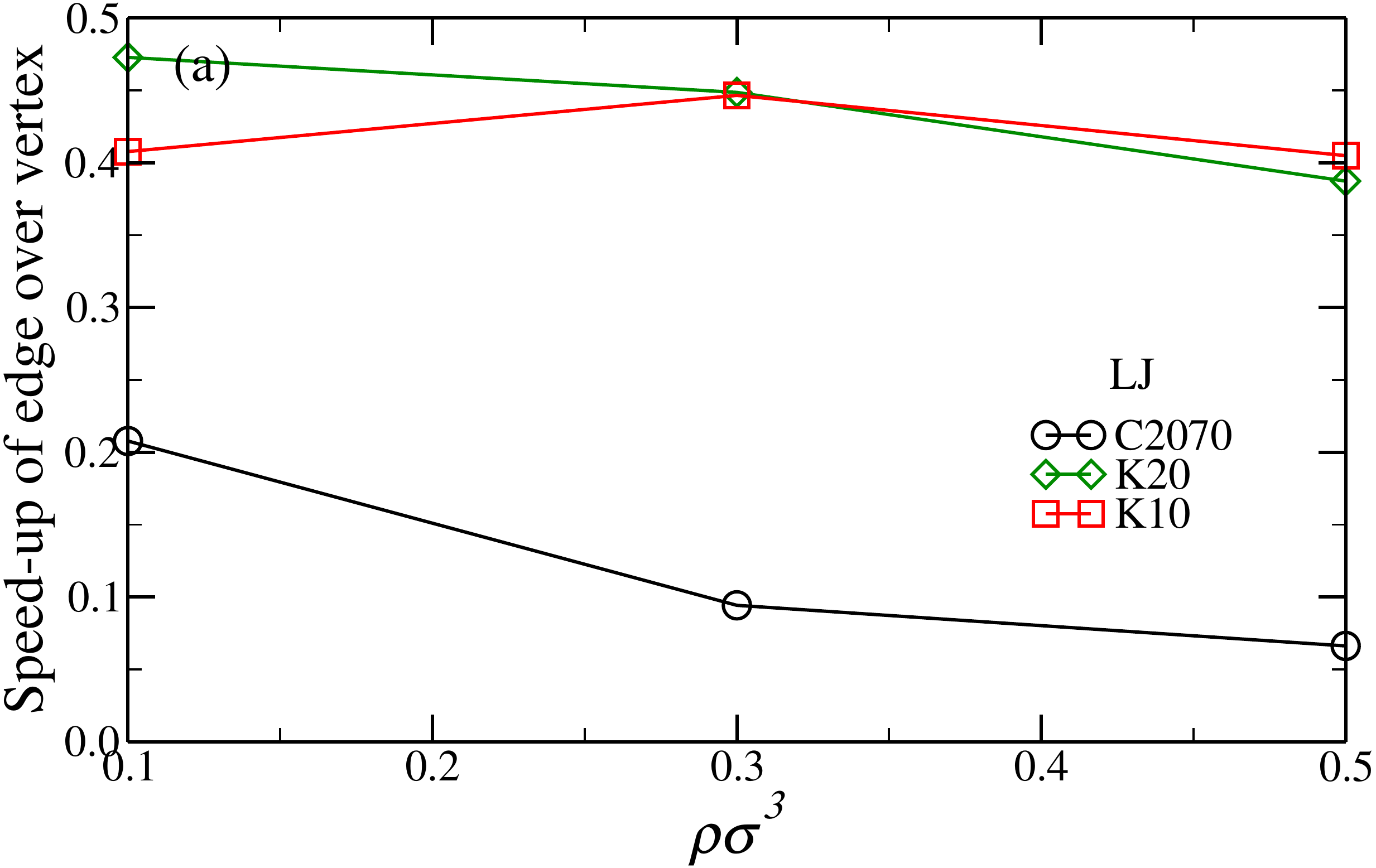}\\ ~\\
    \includegraphics[width=0.45\textwidth]{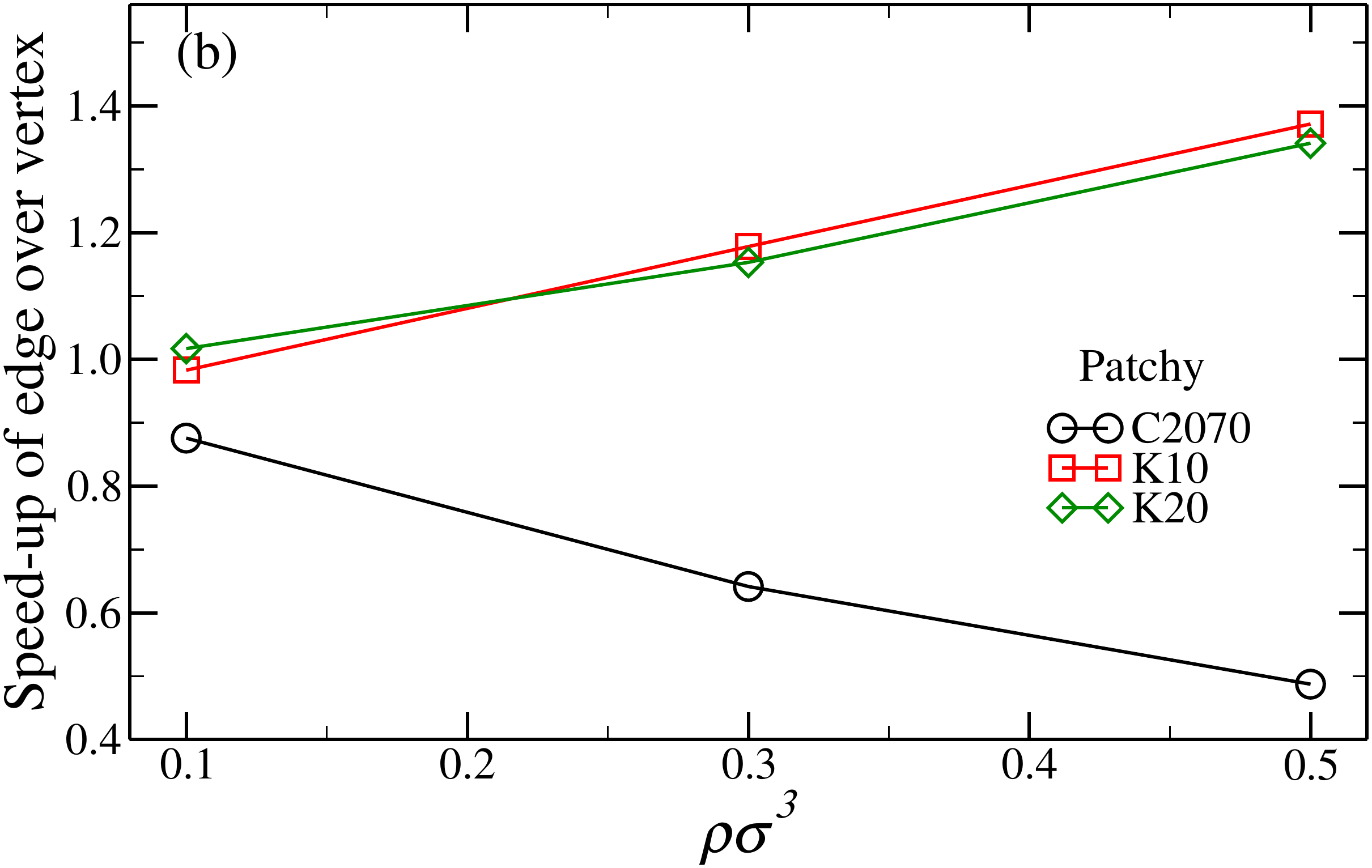}\\
  \end{center}
  \caption{\label{fig:timevsrho}
Achieved speed-up for the edge approach over the vertex approach as a function
of density for (a) the LJ model and (b) the patchy model for all considered GPUs. 
$k_B T/\epsilon = 1.8$ for the LJ model and $0.15$ for the patchy model.
The edge (vertex) approach is more efficient when y-values are larger (smaller) than one.}
\end{figure}

We start the discussion of our results by comparing the performances of the two
approaches for the three different models under standard thermodynamic
conditions. We repeat our measurements for three different GPU architectures,
since these are under active development and tests might give substantially
different results. We run our tests on the NVIDIA cards C2070 (Tesla
architecture, released in 2010), K10 and K20 (Kepler architecture, released in
2012). In Fig.~\ref{fig:timevsN} we report the speed-up due to the edge-based
approach versus the vertex-based approach for the patchy and LJ models at the
highest density $\rho\sigma^3 = 0.5$ and for oxDNA at the only investigated
density. The speed-up is defined as the average running time for the
vertex-based algorithm divided by the average running time of the edge-based
algorithm. The averages are taken over the same number of cycles with the two
algorithms on the same hardware. Each amount of cycles is larger than the
decorrelation time of each system, thus giving timings that are reflective of
its physical properties. We point out that newer hardware has better
performance than old hardware in all cases if the same algorithm and model are
employed. The timings are repeated 5 times for each data point to accumulate
averages.
Since floating point operations are not commutative, and the order of
operations cannot be controlled in an efficient parallel algorithm,
simulations on GPUs (as well as parallel simulations on CPUs) are not
reproducible. We thus have to let each simulation undertake a different
trajectory and make sure that our timings are taken over long enough intervals
to average out the small differences in running time due to the different
sequences of calculations.

All panels in Fig.~\ref{fig:timevsN} show a plateau in the speed-up for a large
enough system, and the height of this plateau is the benchmark that we use to
asses the performance of the algorithms. Since GPUs become faster than a single
CPU core only at large ($N \gtrsim 500-1000$) system sizes, we assume that it
is most relevant to compare GPU codes for large systems. In the case of the LJ
potential, shown in Fig.~\ref{fig:timevsN}(a), the vertex-based approach is the
fastest for all GPUs considered, although newer architectures suffer less from
the introduction of the edge-based approach. This is because the edge-based
approach is slowed down by carrying out many atomic operations, which are
faster in the newest architecture but not yet fast enough to make it
competitive. The patchy model displays a different behaviour
(Fig.~\ref{fig:timevsN}(b)): the edge approach becomes favourable on the most
recent architecture by almost 40\%, compared to a slowdown of 50\% on the
oldest hardware.
It is the substantially improved speed of atomic operations on the newer
hardware architecture that favours the edge-based approach.~\cite{nvidia_kepler_2012} 
The amount of pair
interaction computations is essentially controlled by the density, while the
number of non-zero interactions has a maximum which is dictated by the
interaction potential. This means that the number of interactions that need to
be computed and the number of interactions that are actually non-zero both
increase with density, but the former increases faster than the latter. Slower
atomic operations and a smaller amount of compute units on the older
architecture favour the vertex-based approach, that requires fewer threads and
no atomic operations. The edge-based approach is faster on the newer hardware
because it can better exploit the larger number of threads and is slowed down
less by the atomic operations which it requires.

Finally, the compute-intensive pair interactions of the oxDNA model
(Fig.~\ref{fig:timevsN}(c)) always benefit from the edge-based approach, which
yields a $70\%$ performance increase or more with the newest K20
hardware.~\cite{nvidia_kepler_2012} There are several reasons for this, the
main one being that the vertex approach has a poorer balance between the
workload each thread has to carry out as opposed to the edge-based approach,
which is itself not very well balanced but hides the load imbalance with a much
larger amounts of threads. It is worth pointing out that in general the more
recent the hardware the bigger the speed-up of the edge-based approach, because 
it benefits from more compute units and faster atomic operations.~\cite{nvidia_kepler_2012}

The relative effectiveness of the two parallelization approaches depends on the
amount of potentially interacting neighbours. For the LJ and patchy models this
number can change significantly as the density of the system is changed, and we
thus repeat our tests at three different values of the density. In the DNA
model, since the local environment of the molecules stays the same, the density
does not change significantly the number of potential interactions and hence
performances are very weakly density-dependent, at least in the density range
usually considered in DNA applications. The density dependence of the speed-up
is shown in Fig.~\ref{fig:timevsrho} for the three models. In the case of LJ,
the relative performances stay more or less constant, except for the oldest
hardware where the increased number of atomic operations has a negative impact
on the edge-based approach. In the case of the patchy system, on the other
hand, increasing the density favours the edge-based approach if using new
hardware and favours the vertex-based approach on the older hardware. This can
be rationalized as follows: the number of potential interactions increases
faster with increasing density than the number of non-zero interactions. The
edge-based approach is effective in treating potential interactions that turn
out to be zero, because they do not produce atomic operations. But the latter
nevertheless increase with density, and when using the oldest hardware the
balance is still reversed in favour of the vertex-based approach.

We note that the vertex-based approach reported in this section does not use
Newton's third law. We also measured the performance of the vertex-based
approach which uses Newton's third law, as outlined in Sec.~\ref{sec_methods}.
We found its performance to be inferior for LJ systems in comparison with
vertex-based approach without Newton's third law on all considered
architectures. For the patchy particle systems, we found that on the Kepler
architectures K10 and K20 the vertex-based approach with Newton's third law can
be faster by up to a factor $1.2$, but it is always slower than the edge-based
approach. For the DNA systems, we found the vertex-based approach with Newton's
third law to be slightly faster than the approach without Newton's third law on
both Tesla and Kepler architectures. However, the achieved speed-up was only at
most 16\%, significantly smaller than the speed-up achieved
with the edge-based approach. For completeness, we provide the comparisons of the
vertex-based approaches with and without Newton's third law in Supplementary
Information (Fig. S2), along with the absolute time per MD step for the results 
shown in Figure~\ref{fig:timevsN}.~\cite{support}

\section{Conclusions}

Being able to exploit the impressive computer power of GPUs can be important in
molecular simulations, since these computing devices have the power to study
system sizes and time scales previously untreatable. Unfortunately, this comes
at the cost of rethinking the structure of the simulation codes, since
approaches that are known to fail in CPU programming can turn out to be
effective or vice-versa. We have compared the performances of two different
parallelization approaches, ``vertex''- and ``edge''-based, by simulating three
different models with quite distinct computational complexities. The 
edge-based parallelization approach, where a thread is started for
each potentially non-zero interaction, is competitive and often outperforms the
vertex-based approach, where a thread is started for each particle
in the system. The reason for this is that sometimes the vertex-based approach
is not parallel enough to take full advantage of the GPU.

A vertex-based approach is still the fastest when dealing with very simple
potentials with a relatively large number of neighbours, which is the case when
the interaction range is large and the interaction is spherical. The edge-based
approach is the fastest when the non-zero interactions per particle are few
and/or complicated, giving its best performance when the interaction potentials
are both highly anisotropic and complicated as it is the case in the oxDNA
model. Not surprisingly, the edge-based approach benefits more from an
increased number of scalar processors on the graphics card, which appears to be
the current trend in improving this kind of hardware. We thus predict that the
edge-based parallelization will become more and more competitive in the future,
if the current trends in hardware improvements are continued.

\section*{Acknowledgements}

The authors thank M. Sega for technical help, the Advanced Research Computing, 
University of Oxford for computer time and NVIDIA for the hardware donations. 
F.R. acknowledges financial support from the Engineering and Physical Sciences 
Research Council. P.{\v S}. is grateful for the Bobby Berman and Scatcherd 
European Scholarship awards. L.R. thanks the Physical \& Theoretical Chemistry 
Laboratory, University of Oxford for its hospitality and acknowledges support 
from ERC-226207-PATCHYCOLLOIDS.

\bibliography{edge}

\section*{Supplementary Information}
\vspace{1cm}
\noindent
In this Supplementary information we report a comparison between oxDNA and
HOOMD-blue-blue and LAMMPS, as well as absolute timings for the test cases employed
in the text.
\vspace{1cm}

\section{Comparison to HOOMD-blue and LAMMPS}

\begin{figure}[h!]
 \centering
\includegraphics[width=0.45\textwidth]{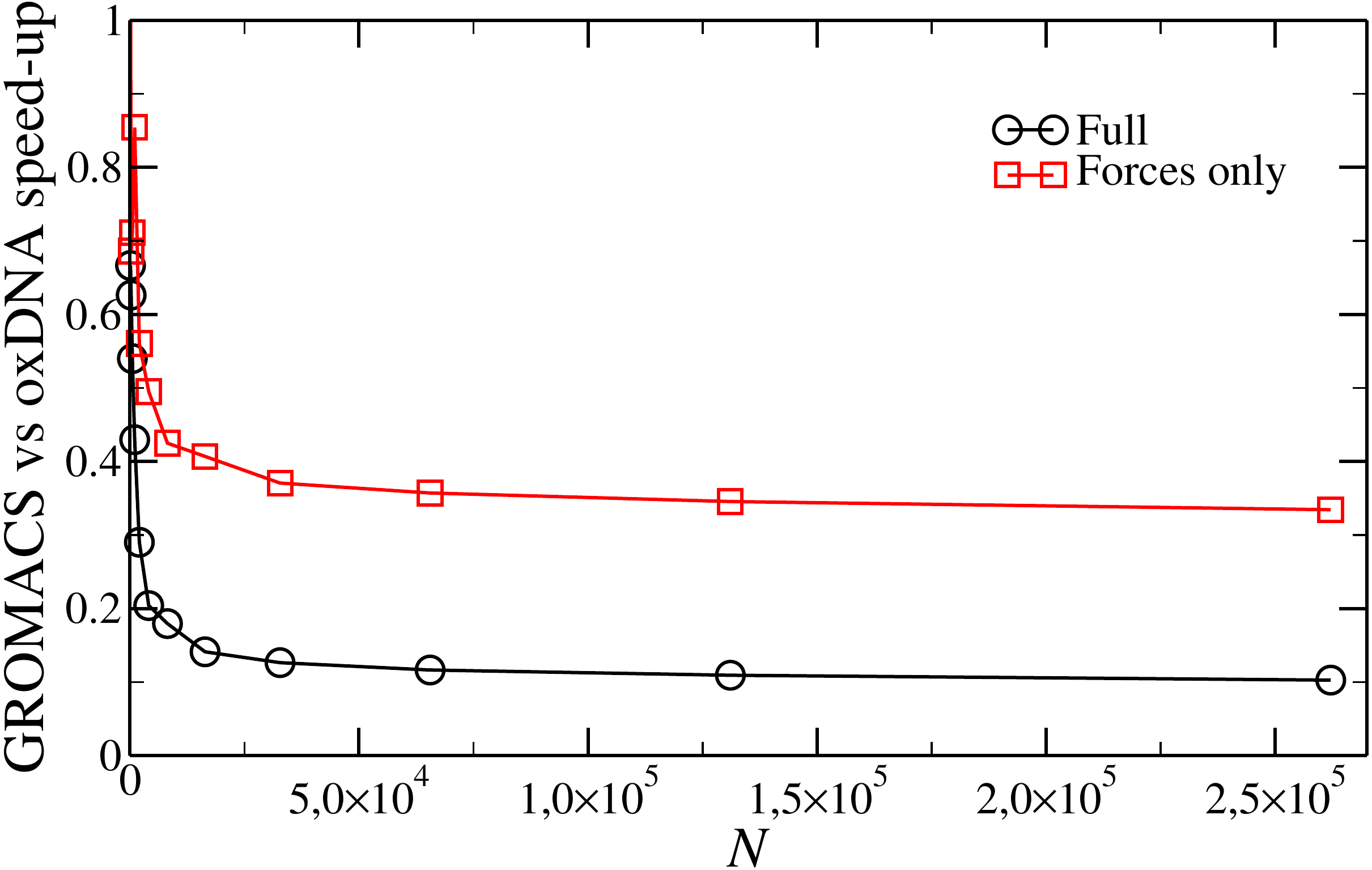}
 \caption{Performance comparison between oxDNA and GROMACS 5.0.1. 
  Timings are carried out on a C2070 NVIDIA GPU. We show the ratio
  between the average time required by oxDNA to perform
 an MD step over the time taken by GROMACS (in black). We also
 show the ratio between the average time required by the kernel
 computing the forces of oxDNA and of GROMACS (in red).}
 \label{fig:comp}
\end{figure}

Since our aim is to compare parallelization algorithms, it is important to
verify that the performances of our code are comparable to state-of-the-art
codes that are publicly available. Figure 2 in the main text shows the comparison
between oxDNA and the GPU-enabled versions of HOOMD-blue (0.11-3 version) and LAMMPS 
(1Feb14 version) for a standard NVE Lennard-Jones simulation at $\rho=0.1$ (vs HOOMD-blue) and 
$\rho=0.8442$ (vs LAMMPS) for several system sizes. Overall, the large-$N$ performances 
are roughly equivalent, with the largest difference being less than 20\%. In addition, the
asymptotic performances of the three codes are very similar.

Figure~\ref{fig:comp} shows a comparison (carried out at $\rho = 0.1$) between oxDNA
and GROMACS 5.0.1. oxDNA is faster for all the considered number of particles. We
stress that GROMACS uses the GPU only to calculate non-bonded forces and thus we 
also show the relative performance between the force-computing kernels in oxDNA
and GROMACS. We note that GROMACS is not tailored to simulate short-ranged-only
potentials and hence it is perhaps unsurprising that it does not perform very well
in this particular test.

\section{Absolute timings and comparison of the vertex-based approach with and without Newton's third law}

\begin{figure}[h!]
 \centering
 \includegraphics[width=0.45\textwidth]{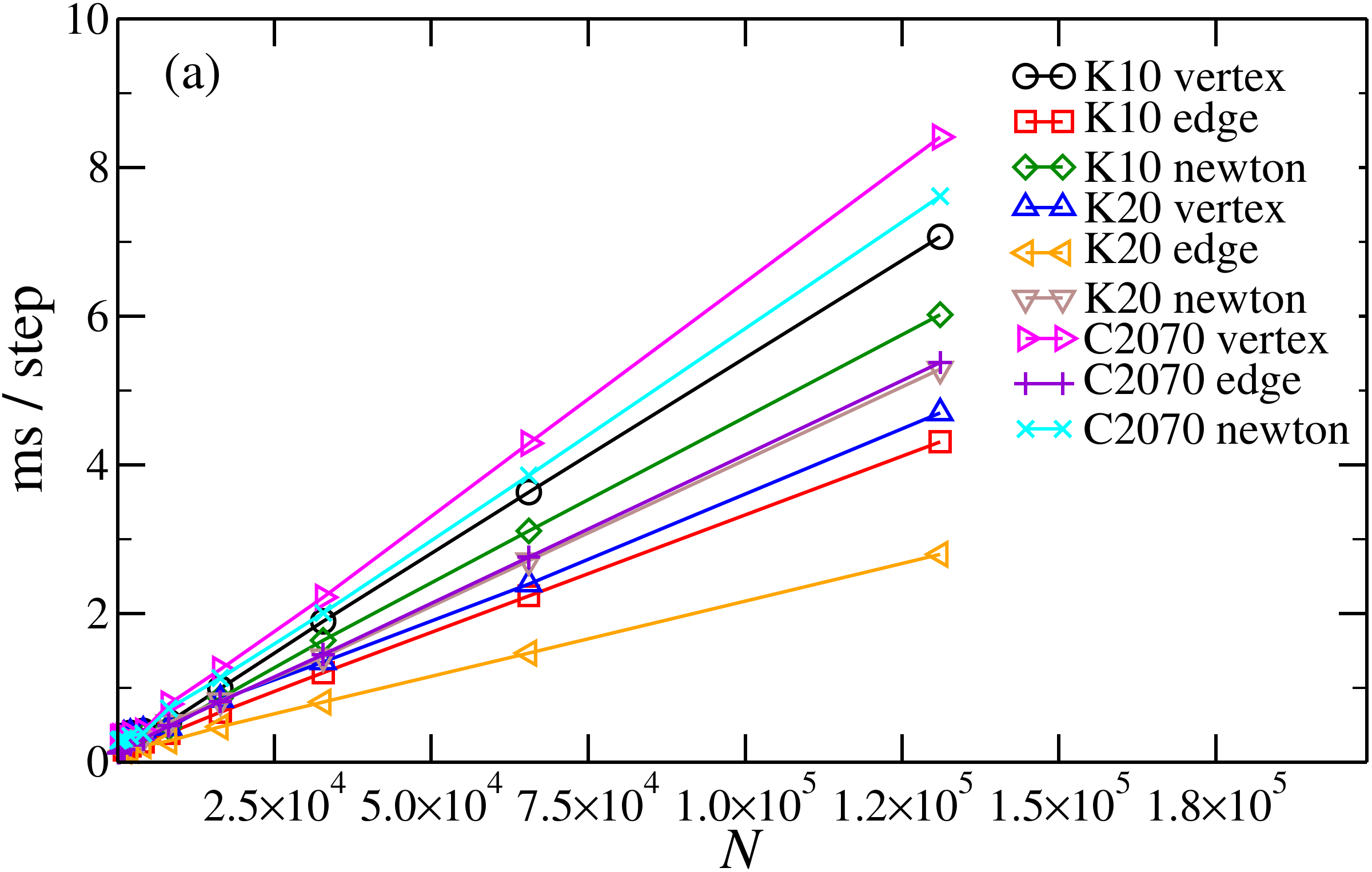}\\ ~\\
 \includegraphics[width=0.45\textwidth]{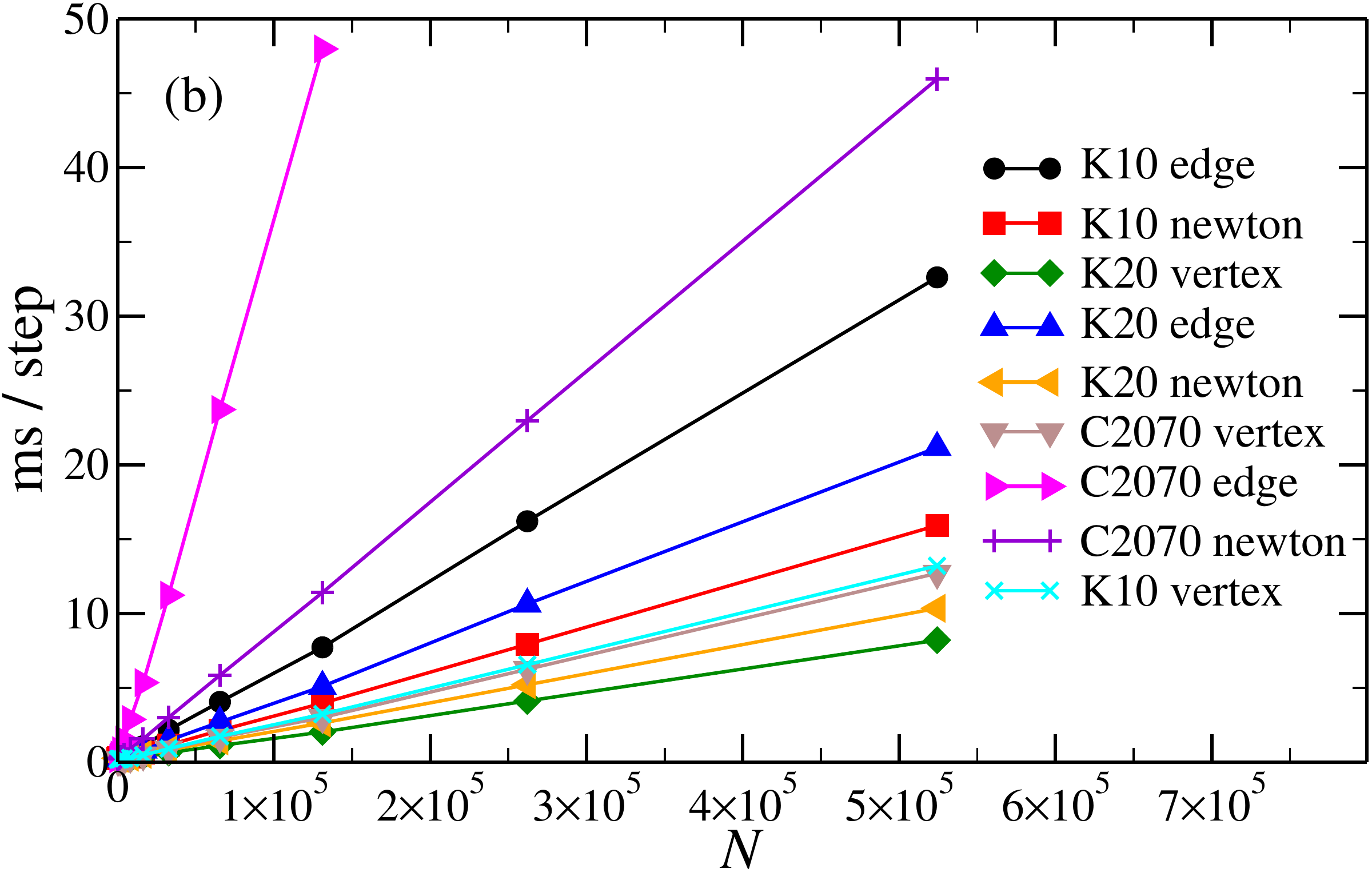}\\ ~\\
 \includegraphics[width=0.45\textwidth]{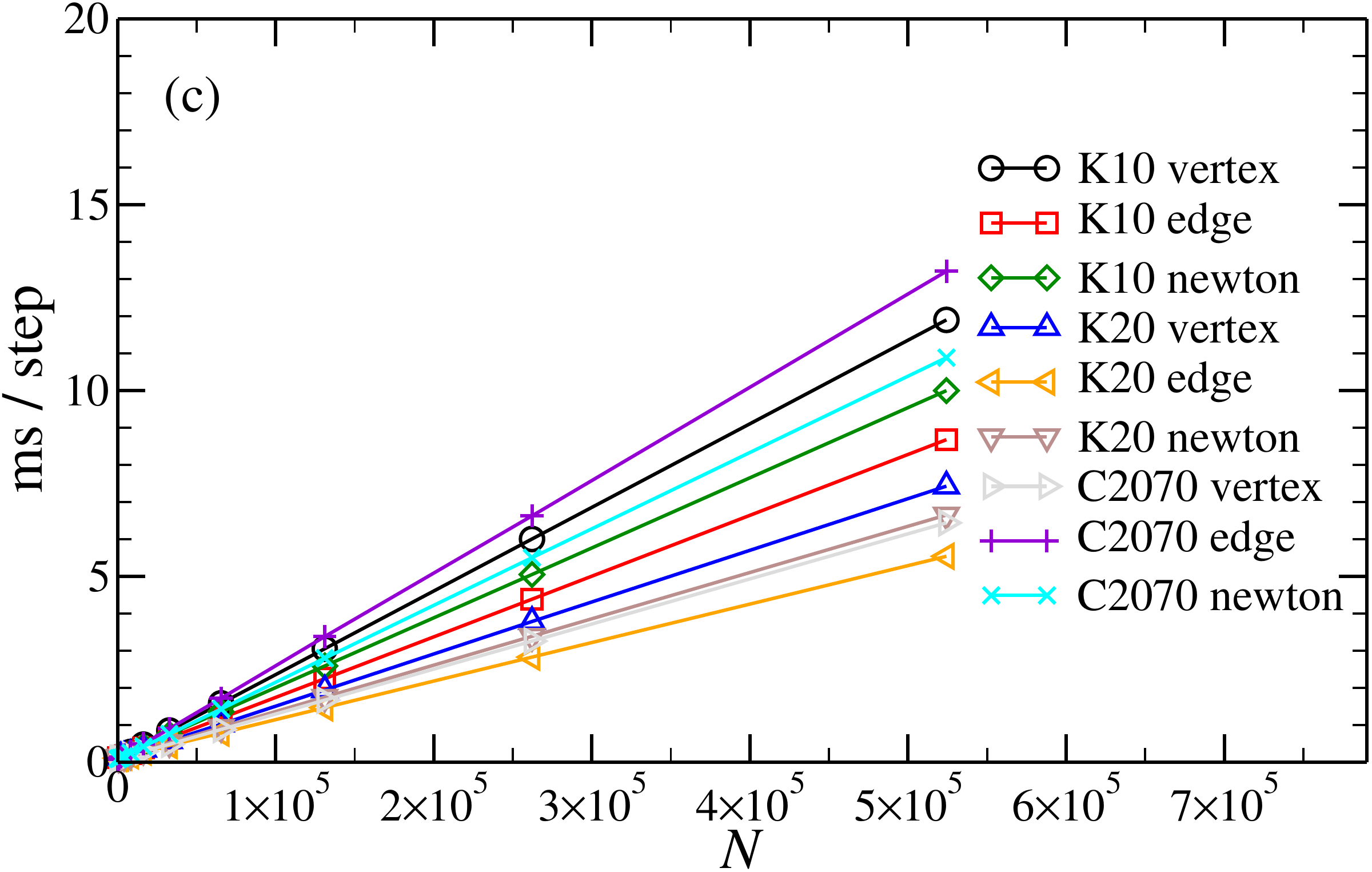}
 \caption{Absolute timings as a function of system sizes. (a) DNA model at a double strand concentration of $2.7\,$mM, $T=300\,$K. (b) 
 LJ model, $k_BT/\epsilon = 1.8$, $\rho\sigma^3 = 0.5$. (c) patchy model, $k_BT/\epsilon = 0.15$, $k_BT/\epsilon = 1.8$, $\rho\sigma^3 = 0.5$}
 \label{fig:timings}
\end{figure}

For reference, we report in Fig.~\ref{fig:timings} the absolute time that is
required for each of the algorithms to perform a single molecular dynamics
step with the different architectures and models. 
The figures include timings for edge-based approach (``edge"), vertex-based 
approach (``vertex") and vertex-based approach with Newton's third law implemented 
(``newton") for C2070, K10 and K20 architectures

\end{document}